\documentclass[aps,prb,showpacs,twocolumn,floats,epsfig]{revtex4}

\usepackage{amsmath}
\usepackage{epsfig}
\usepackage{graphicx}
\usepackage{amsfonts}
\usepackage{color}

\newcommand{\ig}{\includegraphics}
\newcommand{\ct}{\cite}
\newcommand{\bi}{\bibitem}
\newcommand{\be}{\begin{equation}}
\newcommand{\ee}{\end{equation}}
\newcommand{\ba}{\begin{eqnarray}}
\newcommand{\ea}{\end{eqnarray}}
\newcommand{\al}{\alpha}

\newcommand{\ga}{\gamma}
\newcommand{\de}{\delta}
\newcommand{\la}{\lambda}
\newcommand{\si}{\sigma}
\newcommand{\ta}{\theta}
\newcommand{\dg}{\dagger}
\newcommand{\non}{\nonumber}

\newcommand{\bra}{\langle}
\newcommand{\ket}{\rangle}

\begin{document}

\title{Fidelity susceptibility of one-dimensional models with twisted
boundary conditions}
\author{Manisha Thakurathi$^1$, Diptiman Sen$^1$, and Amit Dutta$^2$}
\affiliation{$^1$Center for High Energy Physics, Indian Institute of Science, 
Bangalore 560 012, India \\
$^2$Department of Physics, Indian Institute of Technology, Kanpur 208 016, 
India}

\begin{abstract}
Recently it has been shown that the fidelity of the ground state of a 
quantum many-body system can be used to detect its quantum critical points 
(QCPs). If $g$ denotes the parameter in the Hamiltonian with respect to which 
the fidelity is computed, we find that for one-dimensional models with large 
but finite size, the fidelity susceptibility $\chi_F$ can detect a QCP 
provided that the correlation length exponent satisfies $\nu < 2$. We then 
show that $\chi_F$ can be used to locate a QCP even if $\nu \ge 2$ if we 
introduce boundary conditions labeled by a twist angle $N\ta$, where $N$ is 
the system size. If the QCP lies at $g=0$, we find that if $N$ is kept 
constant, $\chi_F$ has a scaling form given by $\chi_F \sim \ta^{-2/\nu} 
f(g/\ta^{1/\nu})$ if $\ta \ll 2\pi/N$. We illustrate this both in a 
tight-binding model of fermions with a spatially varying chemical potential 
with amplitude $h$ and period $2q$ in which $\nu = q$, and in a $XY$ spin-1/2 
chain in which $\nu =2$. Finally we show that when $q$ is very large, the 
model has two additional QCPs at $h=\pm 2$ which cannot be detected by 
studying the energy spectrum but are clearly detected by $\chi_F$. The peak 
value and width of $\chi_F$ seem to scale as non-trivial powers of $q$ at 
these QCPs. We argue that these QCPs mark a transition between extended and 
localized states at the Fermi energy.
\end{abstract}


\maketitle

\section{Introduction}

Quantum phase transitions have been studied extensively for several years. 
These are transitions which occur in the ground state of a many-body quantum 
system as a parameter $g$ in the Hamiltonian is varied across a critical
value \ct{chakrabarti96, sachdev99,continentino,sondhi97,vojta03}; the ground
state typically has different orders on the two sides of the critical point. 
Several measures arising in quantum information theory, such as entanglement 
\ct{osterloh02,osborne02,roncaglia06}, entanglement entropy 
\ct{vidal03,kitaev06}, 
Loschmidt echo \ct{song06}, decoherence \ct{damski10} and quantum discord 
\ct{olliver01,dillenschneider08,luo08,sarandy09,nag11} have been used to 
detect the location of a quantum critical point (QCP). A number of reviews 
have appeared on the connections between quantum critical systems and quantum 
information theory \ct{amico08,latorre09,polkovnikov,dutta10}.
 
The concept of quantum fidelity ($F$) has proved to be particularly useful for
detecting the locations of QCPs \ct{zanardi06,venuti07,giorda07,buon07,shigu,
schwandt09,gritsev09,zhou,znidaric03,ma08,eriksson09,grandi10,rams11,sirker,
thesberg,mukherjee}. We consider a one-dimensional system 
whose Hamiltonian $H(g)$ contains a parameter $g$ such that the system is at 
a QCP when $g=0$. Note that a QCP exists only in the thermodynamic limit in 
which the number of sites $N \to \infty$.
Consider now a system in which $N$ is finite but large ($\gg 1$). Given two 
ground state wave functions $|\psi_0(g+dg/2)\ket$ and $|\psi_0 (g-dg/2) \ket$ 
at two values of the parameter which are separated by a small amount $dg$ (we 
assume that the ground states are non-degenerate), we define the fidelity as
\be F(g,dg) ~=~ |\bra \psi_0 (g+dg/2)| \psi_0 (g-dg/2)\ket|. \label{fid} \ee
First order perturbation theory shows that in the limit $dg \to 0$, $F -1
\to 0$ as $(dg)^2$. We therefore define the fidelity susceptibility (FS) as 
\be \chi_F (g) ~=~ \lim_{dg \to 0} ~\frac{\ln F(g,dg)}{(dg)^2}. \label{fs1} \ee
(Note that our definition of the FS differs by a factor of 2 from the one
given in many other papers \ct{grandi10,rams11}).
The FS measures how rapidly the ground state changes with $g$. It turns out
that near the QCP at $g=0$, $\chi_F$ shows a large peak even for finite values
of $N$ because there are a large number of very low-energy states which mix 
with each other in a way which changes rapidly with $g$. Thus the FS is able 
to detect the ground state singularities associated with a quantum phase 
transition without making explicit reference to an order parameter. In the 
limit $g \to 0$, it is found that \ct{venuti07,schwandt09,grandi10,gritsev09}
\be \chi_F (g) ~\sim~ - N |g|^{\nu -2}, \label{sca1} \ee
where $\nu$ is the correlation length exponent at the QCP (namely, for 
$N \to \infty$, the correlation length $\xi$ diverges as $|g|^{-\nu}$ as 
$g \to 0$). Eq.~(\ref{sca1}) holds only if $\nu < 2$; the divergence in
that expression for small $g$ arises from contributions from the low-energy
(critical) modes. However, if $\nu \ge 2$, Eq.~(\ref{sca1}) is no longer 
useful for finding QCPs because, for small $g$, the contributions to $\chi_F$ 
from high-energy modes are of the same order as (if $\nu =2$) or dominate over 
(if $\nu > 2$) the terms of order $|g|^{\nu -2}$.

We will show in this paper that even if $\nu \ge 2$, we can use $\chi_F$
to locate a QCP by using twisted boundary conditions labeled by a twist
angle $N\ta$ and taking the limit $\ta \to 0$ in a particular way.
The introduction of $\ta$ allows us to bring the energy of one particular
state arbitrarily close to zero; this state then contributes 
a term to $\chi_F$ which scales with $\ta$ as $\ta^{-2/\nu} f(g/\ta^{1/\nu})$.
In the limit of $\ta \to 0$, a plot of $\chi_F$ versus $g$ clearly shows the 
divergence due to $\ta^{-2/\nu}$ at $g \sim \ta^{1/\nu}$, thereby pinpointing 
the location of the QCP.

The paper is organized as follows. In Sec. II A, we will present a simple 
argument to show that the presence of a twist angle $N\ta$, along with a gap
proportional to $g^\nu$, gives rise to a $\chi_F$ which exhibits the above 
scaling form. In Sec. II B, we indicate how this result may be generalized
to higher dimensions. In Secs. III A-C, we will illustrate how this works in 
a tight-binding model of noninteracting spinless fermions with a periodic 
chemical potential with amplitude $h$ and period $2q$, where $q$ is an 
integer. This model has a QCP at $h=0$ where $\nu = q$; it therefore allows 
us to study the scaling of $\chi_F$ for different values of $\nu$. In Sec. 
III D, we will illustrate the scaling of $\chi_F$ with $\ta$ in a spin-1/2 
chain which has a QCP with $\nu = 2$. In Sec. IV, we will consider the case 
of large values of $q$ and numerically show that $\chi_F$ has a scaling form 
near $h = \pm 2$ with non-trivial power laws. (This study is motivated by the 
observation that the Aubry-Andre model, which has a quasiperiodic chemical 
potential, has transition from extended to localized wave functions at $h = 
\pm 2$). In Sec. V, we will summarize the results presented in this paper.

\section{Fidelity susceptibility for a two-state system}

In this section we will do a simple calculation to determine the fidelity of 
a two-state system. The results obtained here will be used in the following 
sections.

\subsection{One-dimensional systems}

Consider a two-state problem which is governed by a Hamiltonian of the form
\be H ~=~ \left( \begin{array}{cc}
-uk & ag^\nu \\
ag^\nu & uk \end{array} \right). \label{two} \ee
This form is motivated as follows. In later sections we will consider a 
one-dimensional model in which $\pm k$ will denote the momenta of two states 
close to zero energy (measured with respect to the Fermi energy), $u$ will 
denote the Fermi velocity obtained by linearizing the dispersion near zero 
energy, $g$ will denote a parameter in a many-body Hamiltonian such that the 
QCP lies at $g=0$, and $a$ is some constant. The form in Eq.~(\ref{two}) 
will be taken to be valid only for 
values of $k$ much smaller than some cut-off, i.e., only for low-energy modes.
Since the eigenvalues of $H$ in Eq.~(\ref{two}) are equal to $\pm 
\sqrt{u^2 k^2 + a^2 g^{2\nu}}$, we see that the excitation spectrum is given by
$E = \pm u k$ at the QCP (which would be consistent with the dynamical critical 
exponent of a many-body system being given by $z=1$), while $E = g^\nu$ for 
$k=0$ is consistent, for $z=1$, with the definition of the correlation length 
exponent $\nu$. 

After finding the ground state $|\psi_0 (g)\ket$ of $H$ in Eq.~(\ref{two}), 
we can compute the FS using Eqs.~(\ref{fid}-\ref{fs1}); we find
\be \chi_F (g,k) ~=~ - ~\frac{\nu^2}{8g^2} ~\frac{1}{[ag^\nu /(uk) ~+~ 
uk/(ag^\nu)]^2}. \label{fs2} \ee
In the limit $k \to 0$, we see that the FS has the scaling form
\ba \chi_F (g,k) &=& (uk/a)^{-2/\nu} ~f(a^{1/\nu} g/(uk)^{1/\nu}), \non \\
{\rm where}~~~ f(x) &=& - ~\frac{\nu^2}{8x^2} ~\frac{1}{(x^\nu ~+~ 
1/x^\nu)^2}. \label{scal1} \ea
We note that $f(x)$ has a single peak at $x=0$ if $\nu =1$. If $\nu > 1$,
$f(x)$ has two peaks at $x = \pm [(\nu -1)/(\nu +1)]^{1/(2\nu)}$, and goes to
0 as $-(\nu^2/8) x^{2(\nu -1)}$ for $x \to 0$ and as $-\nu^2/(8x^{2(\nu 
+1)})$ for $x \to \infty$.

Next, let us suppose that we are considering a many-body system with size $N$,
so that the momenta are quantized as $k=\ta + 2\pi n/N$, where $n$ 
runs over all integers, and $\ta$ comes from a twisted boundary condition 
to be elaborated in later sections. The presence of the $2\pi n/N$ term 
implies that we can take $\ta$ to lie in the range $[0,2\pi/N]$. If the 
Hamiltonian of the system can be decomposed into independent two-state systems
of the form given in Eq.~(\ref{two}), the FS will be given by
\ba \chi_F (g) &=& - \frac{\nu^2}{8g^2} ~\sum_{n=-\infty}^\infty ~\frac{1}{
\left( \frac{ag^\nu}{u(\ta +2\pi n/N)}~+~ \frac{u(\ta +2\pi n/N)}{ag^\nu} 
\right)^2}. \non \\
& & \label{fs3} \ea
We can now consider the value of $\chi_F$ in various limits.

First, if $N$ is held fixed and we consider a regime in which $ag^\nu, ~u\ta
\ll 2\pi u/N$ and $a^{1/\nu}g/(u\ta)^{1/\nu}$ is of order 1, the sum in 
Eq.~(\ref{fs3}) 
will be dominated by the $n=0$ term and we will obtain the scaling form
\be \chi_F (g) ~=~ (u\ta/a)^{-2/\nu} ~f(a^{1/\nu} g/(u\ta)^{1/\nu}), 
\label{scal2} \ee
where the function $f$ is given in Eq.~(\ref{scal1}). Secondly, if 
$g^\nu \ll 2\pi u/N$ but $\ta$ is of order $2\pi/N$, then we obtain
\ba \chi_F (g) &=& - ~\frac{\nu^2 a^2 g^{2(\nu -1)}}{8u^2} ~
\sum_{n=-\infty}^\infty ~\frac{1}{(\ta +2\pi n/N)^2} \non \\
&=& - ~\frac{N^2 a^2 \nu^2 g^{2(\nu -1)}}{32 u^2 \sin^2 (N\ta/2)}. 
\label{fs4} \ea
As $g \to 0$, Eq.~(\ref{fs4}) shows that $\chi_F$ has a finite limit if 
$\nu =1$ but it goes to 0 if $\nu > 1$. [For the models that we will consider 
later, Eq.~(\ref{two}) is {\it not} a good approximation for $|k| \gg g^\nu$ 
(high-energy modes), and the contributions to the FS from such modes generally
approaches a non-zero value as $g \to 0$ as we will show in 
Eq.~(\ref{fs8}) for $\nu > 1$]. Finally, if $2\pi/N \ll g^\nu$,
the sum in Eq.~(\ref{fs3}) can be replaced by an integral over
$k=2\pi n/N$; if we further assume that $g^\nu \ll 1$, the limits of the
integral can be taken to be $\pm \infty$ since the contributions from
the regions with $|k| \gg g^\nu$ will be negligible. We then obtain
\ba \chi_F (g) &=& - ~\frac{N \nu^2}{8g^2} ~\int_{-\infty}^\infty ~
\frac{dk}{2\pi}~ \frac{1}{[ag^\nu /(uk) ~+~ uk/(ag^\nu)]^2} \non \\
&=& - \frac{N \nu^2 a g^{\nu-2}}{32u}. \label{fs5} \ea
[The integral in the above equation only gets substantial contributions from 
values of $|k|$ lying around $g^\nu$, i.e., from low-energy modes. The 
scaling given in Eq.~(\ref{fs5}) will therefore be valid as long as 
Eq.~(\ref{two}) is a good approximation for such low-energy modes, even
if it fails for high-energy modes]. We thus see from 
Eq.~(\ref{fs5}) that as $g \to 0$, the FS will diverge if $\nu < 2$ but not 
if $\nu \ge 2$. A plot of the FS versus $g$ will therefore show a divergence 
at the QCP ($g=0$) only if $\nu < 2$, but for $\nu \ge 2$, the FS will not be 
useful for finding the location of the QCP.

We conclude that for a large but finite value of the system size $N$, the FS 
can detect a QCP only if $\nu < 2$. Note that this conclusion would not 
change if we replaced $k$ and $g^\nu$ by $k^z$ and $g^{z\nu}$ respectively in 
Eq.~(\ref{two}), with a dynamical critical exponent $z$ which is not equal to 
1. The exponent which appears in the second line in Eq.~(\ref{fs5}) will 
be $\nu -2$ regardless of the value of $z$.

\subsection{Higher dimensional systems}

Although the rest of this paper will be only concerned with one-dimensional
systems, let us briefly discuss what may happen in higher dimensions. We 
assume that there is a $d$-dimensional system in which the modes with momenta 
$\pm {\vec k}$ are governed by a Hamiltonian of the form given in 
Eq.~(\ref{two}), where the momentum variable in that equation now stands for 
$k= |\vec k|$. Eqs.~(\ref{fs2}-\ref{scal1}) will continue to hold. Next,
let us impose periodic boundary conditions, with a twist angle $\ta$ in one of 
the $d$ dimensions and zero twist angle in the remaining $d-1$ dimensions. 
To be specific, we assume that we 
have a hypercubic lattice system which has $N$ sites in each dimension, so 
that the different momenta are quantized as $k_i = 2 \pi n_i /N$; the twist 
shifts the first momentum to $k_1 = \ta + 2 \pi n_1 /N$. Eq.~(\ref{fs3}) then 
gets modified to 
\ba \chi_F (g) &=& - \frac{\nu^2}{8g^2} ~\sum_{n_1,\cdots,n_d=
-\infty}^\infty ~\frac{1}{\left( \frac{ag^\nu}{uk'}~+~ \frac{uk'}{ag^\nu} 
\right)^2}, \non \\
k' &=& [ (\ta + 2 \pi n_1 /N)^2 + \sum_{i=2}^d (2 \pi n_i /N)^2 ]^{1/2}. 
\label{fs9} \ea
As before, we can now consider what happens in three different cases.

If $N$ is held fixed, $ag^\nu, ~u\ta \ll 2\pi u/N$ and $g/(u\ta)^{1/\nu}$ is 
of order 1, the sum in Eq.~(\ref{fs9}) will be dominated by the term in which 
$n_i=0$ for all $i$. Then we will obtain the scaling form given in 
Eq.~(\ref{scal2}) where the function $f$ is given in Eq.~(\ref{scal1}). Note 
that the scaling form does not depend on the dimensionality $d$ in this case.
Next, if $g^\nu \ll 2\pi u/N$ but $\ta$ is of order $2\pi/N$, then we obtain
\ba & & \chi_F (g) ~=~ - ~\frac{N^2 \nu^2 a^2 g^{2(\nu -1)}}{8u^2} \non \\
& & \times \sum_{n_1,\cdots,n_d=-\infty}^\infty ~\frac{1}{(N \ta +2\pi 
n_1)^2 + \sum_{i=2}^d (2\pi n_i)^2}. \non \\
&& \label{fs10} \ea
As $g \to 0$, Eq.~(\ref{fs10}) has a finite limit (which depends on $N\ta$)
if $\nu =1$ but goes to 0 if $\nu > 1$. Finally, if $2\pi/N \ll g^\nu$,
the sum in Eq.~(\ref{fs9}) can be replaced by an integral over
$k_i=2\pi n_i/N$; assuming that $g^\nu \ll 1$, the limits of the
integral can be taken to be $\pm \infty$. We then obtain
\ba \chi_F (g) &=& - ~\frac{N^d \nu^2}{8g^2} \int_{-\infty}^\infty 
\frac{d^dk}{(2\pi)^d} \frac{1}{[ag^\nu /(u|\vec k|) + u|\vec k|/(ag^\nu)]^2} 
\non \\ 
&\sim& - ~N^d a^d g^{d\nu-2}. \label{fs11} \ea
(A similar scaling relation was found by adiabatic perturbation theory 
\ct{gritsev09,grandi10}). As $g \to 0$, Eq.~(\ref{fs11}) diverges if $d\nu < 
2$ but not if $d\nu \ge 2$. A plot of $\chi_F$ therefore diverges at the QCP 
if $d\nu < 2$, but for $d\nu \ge 2$, $\chi_F$ is not useful for locating the 
QCP.

\section{Tight-binding model with variable $\nu$}

We will now study fidelity in a one-dimensional tight-binding model of
spinless fermions in which the exponent $\nu$ can take any integer value;
this will illustrate many of the points discussed in Sec. II. In Sec. III A, 
we will discuss the model and its energy spectrum close to zero energy, while
in Sec. III B, we will examine a number of features related to fidelity.

\subsection{The model}

The model that we are interested in was studied recently from the point of 
view of quenching dynamics across a QCP \ct{manisha,smitha}. In this section,
we will summarize the relevant discussion from Ref. \onlinecite{manisha}. 
The Hamiltonian is given by
\ba H &=& - \sum_{n=1}^N ~[~ J ~(e^{i\ta} c_n^\dg c_{n+1} + e^{-i\ta} 
c_{n+1}^\dg c_n) \non \\
& & ~~~~~~~~~~+~ h ~\cos (\frac{\pi n}{q} ~+~ \phi) ~c_n^\dg c_n], 
\label{ham1} \ea
where $q$ is a positive integer, $N$ is the system size, and we have imposed 
periodic boundary conditions so that $c_{N+1} \equiv c_1$. (We will set the 
hopping amplitude $J$, $\hbar$ and the lattice spacing $a$ equal to unity). 
Note that one can perform a unitary transformation on the $c_n$, namely,
$c_n \to e^{-in\ta} c_n$, which removes the phase $\ta$ from all
the hopping terms except for the hopping between the sites at $N$ and 1
where the phase becomes $e^{\pm i N \ta}$; this is called a twisted
boundary condition, with $N\ta$ being the twist angle. We can assume that
$0 \le N \ta \le 2 \pi$, namely, that $\ta$ lies in the range $[0,2\pi/N]$.
The usual periodic boundary condition corresponds to $\ta = 0$.
The period of the chemical potential in Eq.~(\ref{ham1}) is $2q$, and we will 
assume that $N$ is a multiple of $2q$. A chemical potential of this form 
appears in the Azbel-Hofstadter model 
\ct{azbel,hsu,aubry,ober,ost,sok,delyon,sun,wieg,sen,aulbach,modugno}.
Since shifting $n \to n+1$ in Eq.~(\ref{ham1}) is equivalent to shifting 
$\phi \to \phi + \pi/q$, we can assume without loss of generality that 
$\phi$ lies in the range $[0,\pi/q]$.

The fermionic operators can be Fourier transformed to the momentum basis, 
\be c_k ~=~ \frac{1}{\sqrt N} ~\sum_{n=1}^N c_n e^{-ikn}, \ee
where the momentum $k$ goes from $-\pi$ to $\pi$ in units of $2\pi/N$;
these operators satisfy the anti-commutation rules $\{ c_k, c_{k'}^\dg \} 
= \delta_{k,k'}$. In momentum space, the first two terms of the Hamiltonian 
in Eq.~(\ref{ham1}) have the tight-binding form
\be H_0 ~=~ -\sum_{k=-\pi}^\pi ~2\cos (k+\ta) ~c_k^\dg c_k. \label{ham2} \ee 
For the last term in Eq.~(\ref{ham1}), we use the decomposition
\be h \cos (\frac{\pi n}{q} + \phi) ~=~ \frac{h}{2} ~(e^{i(\pi n/q + \phi)} ~
+~ e^{-i(\pi n/q + \phi)}). \ee
Hence this term couples two fermionic modes with momenta $k_1$ and $k_2$
if $k_1 = k_2 \pm \pi /q$. This fragments the total Hamiltonian $H$ into 
$N/(2q)$ decoupled Hamiltonians $H_k$ which are labeled by a momentum $k$ 
lying in the range $-\pi$ to $-\pi + \pi/q$, namely,
\be H ~=~ \sum_{k=-\pi}^{-\pi+\pi/q} ~H_k. \ee
For each $k$, $H_k$ can be written as a $(2q)$-dimensional matrix involving 
the momenta $k + r\pi/q$, where $r=1,2,\cdots,2q$. The matrix elements of 
$H_k$ are given by
\ba && \bra k+r\pi/q | H_k | k+s\pi/q \ket \non \\
&& = -2 \cos(k+\ta+s\pi/q) \de_{r,s} \non \\
&& - (h/2) (e^{i\phi} \de_{r,s+1} + e^{-i\phi} \de_{r,s-1}), \label{hk} \ea
where $1 \le r,s \le 2q$, and we have set $J=1$. In writing Eq.~(\ref{hk}),
we have assumed `periodic boundary conditions' for the matrix $H_k$, so that 
$r=0$ and $2q+1$ mean $r=2q$ and $1$ respectively. 

Before proceeding further, let us point out an interesting consequence of the
decoupling of $H$ into a sum of $H_k$ and then a duality symmetry 
\ct{aubry,sok}. 
Suppose that the system has a parameter $\ta$ and $N=2qr$ sites, where $r$ is 
a positive integer. Since the momentum $k$ is quantized in units of $2\pi/N$, 
and only momenta differing by integer multiples of $\pi/q$ are coupled 
to each other, the form of $H_k$ in Eq.~(\ref{hk}) implies that this
system is exactly equivalent to a sum of $r$ different systems, which 
have $N'=2q$ sites each but have $r$ different values of $\ta'$ given by
$\ta, ~\ta + 2\pi/(2qr), ~\ta + 4\pi/(2qr), ~\cdots, ~\ta + 2(r-1)\pi/(2qr)$.
Therefore, instead of studying a system with $2qr$ sites and one particular
value of $\ta$, we can study a system with only $2q$ sites but several values 
of $\ta$. Next, we observe that if $N=2q$, the Hamiltonians in real and 
momentum space, Eqs.~(\ref{ham1}) and (\ref{hk}), get mapped into each other 
if we simultaneously interchange $J \leftrightarrow h/2$ and $\ta 
\leftrightarrow \phi$; here we have used the fact that a shift in $\phi$ or 
$\ta$ by $\pi/q = 2\pi/N$ has no effect on the spectrum. In particular, the 
model in Eq.~(\ref{ham1}) is self-dual if $J=h/2$ and $\ta = \phi$.

For each of the Hamiltonians $H_k$ in Eq.~(\ref{hk}), the $2q$ energy levels 
come in $q$ pairs $\pm E$. This can be shown by shifting $s \to s+q$ which
flips the sign of the first term in Eq.~(\ref{hk}), and changing the
sign of the state corresponding to $s$ by $(-1)^s$ which flips the sign
of the terms proportional to $h$ in that equation. Assuming that there are
no states at exactly zero energy, we see that the ground state of each
$H_k$ is one in which the $q$ negative energy states are filled and the
$q$ positive energy states are empty; thus the ground state is half-filled
for each $H_k$.

It will soon become clear that the model defined above has a QCP at $h=0$. 
We therefore use perturbation theory to study the region near $h=0$
\ct{manisha}. (We will set $\ta = 0$ in
Eqs.~(\ref{pert1}-\ref{heff}); this is a reasonable approximation if $N$ 
is large since we always take $0 \le \ta \le 2\pi/N$). We will be particularly
interested in states near zero energy which will contribute the most to the 
FS. Since the system is at half-filling, these states lie near 
the momenta $k = \pm k_F$, where the Fermi momentum $k_F = \pi/2$. If the 
amplitude $h$ of the chemical potential is zero, the system is gapless and the 
states at $k = \pm \pi/2$ are degenerate with each other. Assuming that $h$
is small compared to the band width of $4J$, we can use a perturbative 
expansion in $h$ to calculate the breaking of this degeneracy. 
The regions near $k=-\pi/2$ and $k=\pi/2$ regions are coupled 
through one series of intermediate states lying at $k=-\pi/2 + \pi/q$, $-\pi/2
+ 2\pi/q$, $\cdots$, $\pi/2 - \pi/q$ (with an amplitude equal to $-he^{i\phi}$
at each step), and through another series of intermediate states lying at 
$k=-\pi/2 - \pi/q$, $-\pi/2 - 2\pi/q$, $\cdots$, $\pi/2 + \pi/q$ (with an 
amplitude equal to $-he^{-i\phi}$ at each step). Each of these series 
consists of $q-1$ intermediate states. At the $q$-th order in perturbation 
theory, we therefore obtain an effective Hamiltonian $H_{eff}$ which
has a matrix element between the states at $k = \pm \pi/2$ given by
\ba && \Delta \equiv \bra \pi/2 | H_{eff} | - \pi/2 \ket ~=~ \bra - 
\pi/2 | H_{eff} | \pi/2 \ket^* \non \\ 
&& = ~\frac{(-h/2)^q ~e^{iq\phi}}{\prod_{s=1}^{q-1} ~(2\cos(-\pi/2 + s\pi/q))}
\non \\
&& ~+~ \frac{(-h/2)^q ~e^{-iq\phi}}{\prod_{s=1}^{q-1} ~(2\cos(-\pi/2 - 
s\pi/q))}, \label{pert1} \ea 
where the denominators come from factors like $E_{-\pi/2} - E_{-\pi/2 \pm 
s\pi/q} = 2\cos(-\pi/2 + \pm s\pi/q)$ corresponding to the energies in the 
unperturbed Hamiltonian in Eq.~(\ref{ham2}). 
Simplifying Eq.~(\ref{pert1}) gives
\be \Delta ~=~ (-1)^q (h/2)^q ~\frac{e^{iq\phi} + (-1)^{q-1} e^{-iq\phi}}{
\prod_{s=1}^{q-1}~ (2\sin(\pi s/q))}. \label{pert2} \ee
Hence the magnitude of $\Delta$ is given by
\ba |\Delta| &=& \frac{h^q}{4^{q-1}} ~\frac{|\cos (q\phi)|}{\prod_{s=1}^{
q-1}~ \sin(\pi s/q)} ~~~{\rm if}~~ q ~{\rm ~is ~odd}, \non \\
&=& \frac{h^q}{4^{q-1}} ~\frac{|\sin (q\phi)|}{\prod_{s=1}^{q-1}~ \sin
(\pi s/q)}~~~{\rm if}~~ q ~{\rm ~is ~even}. \label{pert3} \ea
We see that $\phi$ governs the relative phase between the two 
sets of intermediate states which connect the states at $k=\pm \pi/2$. 
We will assume that $\phi$ is such that $\cos (q\phi) \ne 0$ if $q$ is
odd, and $\sin (q\phi) \ne 0$ if $q$ is even; if these conditions are
violated, we would have to go to higher order perturbation theory to find
a non-zero matrix element connecting the states at $k=\pm \pi/2$.

We can now consider moving slightly away from $k=\pm \pi/2$; then the 
unperturbed energies of the states $k=-\pi/2 + k'$ and $\pi/2 + k'$ are 
given by $-2 k'$ and $2 k'$ respectively. The effective Hamiltonian
describing these two states is then given by the $2 \times 2$ matrix
\be H_{eff,k'} ~=~ \left( \begin{array}{cc}
- 2 k' & \Delta^* \\
\Delta & 2 k' \end{array} \right), \label{heff} \ee
where we assume that $\Delta$ continues to be given by the expression in 
Eq.~(\ref{pert2}) because $k'$ is small. The eigenvalues of (\ref{heff})
are given by $\pm \sqrt{4 k'^2 + | \Delta|^2}$; this is the dispersion 
of a massive relativistic particle whose velocity is equal to the Fermi 
velocity $u = 2$ and mass is 
proportional to $|\Delta| \sim h^q$ times $\cos (q\phi)$ or $\sin (q\phi)$.

Hence, $h=0$ corresponds to a QCP where the mass
gap vanishes. Given that the energy vanishes as $|k'|$ if $h=0$ and as $h^q$ 
if $k' = 0$, the dynamical critical exponent and correlation length exponent 
are given by $z=1$ and $\nu = q$, respectively. The correlation length 
exponent $\nu$ thus depends in a simple way on the periodicity of the 
chemical potential. 

Comparing Eq. (\ref{two}) with Eqs.~(\ref{pert3}-\ref{heff}), we identify 
$g=h$, $u=2$, and the constant $a$ is given by
\ba a &=& \frac{1}{4^{q-1}} ~\frac{|\cos (q\phi)|}{\prod_{s=1}^{q-1}~ \sin(\pi
s/q)} ~~~{\rm if}~~ q ~{\rm ~is ~odd}, \non \\
&=& \frac{1}{4^{q-1}} ~\frac{|\sin (q\phi)|}{\prod_{s=1}^{q-1}~ \sin
(\pi s/q)}~~~{\rm if}~~ q ~{\rm ~is ~even}. \label{const} \ea

To summarize, we began with a model whose Hamiltonian in momentum space 
consists of a sum of $(2q)$-dimensional Hamiltonians. Close to the QCP which 
lies at $h=0$, we used perturbation theory to write an effective two-state 
Hamiltonian which governs pairs of low-energy states, i.e., states close to 
the Fermi energy. In the next section, we will use the results obtained in 
Sec. II A for the FS of two-state systems to obtain the FS of our model.

\subsection{Fidelity susceptibility for various $\nu$}

We begin this section by describing how the FS can be numerically calculated
for the model presented in Sec. III A. As we have seen, the Hamiltonian
decouples into $N/(2q)$ Hamiltonians $H_k$. If we can compute the fidelity 
$F_k (h,dh) = |\bra \psi_{0,k} (h+dh) | \psi_{0,k} (h) \ket |$ for each of 
the $H_k$, the total fidelity of the system will be given by the product
\be F (h,dh) ~=~ \bigotimes_{k=-\pi}^{-\pi + \pi/q} ~F_k (h,dh). \ee
Next, we know that the ground state is half-filled for each of the $H_k$ for 
every value of $h$. Let $\psi_{i,k} (h)$ denote the first quantized wave 
functions of the filled states $i$ ($i=1,2,\cdots,q$); each of the $\psi_{i,k}
(h)$ denotes a $(2q)$-dimensional column which is the space on which $H_k$ 
acts. Let $\psi_{i,r,k} (h)$ denote the $r$-th component of $\psi_{i,k} (h)$
($r=1,2,\cdots,2q$), and $c_{r,k}^\dg$ denote the second quantized creation 
operator for the fermion corresponding to the $r$-th component. Then the 
$i$-th filled state can be written in second quantized form as 
${\hat \psi}_{i,k}^\dg (h) | vac \ket$, where
\ba {\hat \psi}_{i,k}^\dg (h) &=& \sum_{r=1}^{2q}~ \psi_{i,r,k} (h) ~
c_{r,k}^\dg, \non \\
{\hat \psi}_{i,k} (h) &=& \sum_{r=1}^{2q}~ \psi_{i,r,k}^* (h) ~
c_{r,k}, \ea
and $| vac \ket$ denotes the vacuum state of the fermions: $c_{r,k} | vac \ket
= 0$ for all $r$ and $k$. The half-filled ground state is therefore given by 
the second quantized expression
\be \bigotimes_{i=1}^q~ {\hat \psi}_{i,k}^\dg (h) | vac \ket. \ee
We can now use Wick's theorem \ct{itzykson} to show that
\ba F_k (h,dh) &=& |\bra vac | \bigotimes_{i=1}^q~ {\hat \psi}_{i,k} (h+dh/2) 
\non \\
&& ~~~~~~\bigotimes_{j=1}^q~ {\hat \psi}_{j,k}^\dg (h-dh/2) |vac \ket | \ea
is given by the magnitude of the determinant of a $q$-dimensional matrix 
$M_k (h,dh)$,
\ba F_k (h,dh) &=& |det [M_k(h,dh)]|, \non \\
(M_k (h,dh))_{ij} &=& \bra vac | {\hat \psi}_{i,k} (h+dh/2) {\hat 
\psi}_{j,k}^\dg (h-dh/2) | vac \ket \non \\
&=& \sum_{r=1}^{2q} ~\psi_{i,r,k}^* (h+dh/2) \psi_{j,r,k} (h-dh/2). \non \\
&& \label{det} \ea
Thus the computation of the total fidelity $F$ reduces to finding the 
determinants of the matrices $M_k$ and then multiplying the determinants 
over $N/(2q)$ values of $k$. The FS is then found as
\be \chi_F (h) ~=~ \lim_{dh \to 0} ~\frac{\ln F(h,dh)}{(dh)^2}. \label{fs6} \ee

We can now use the results obtained in Sec. II to understand various properties
of the FS of the model discussed in Sec. III A; we have seen that $\nu =q$ for 
this model. Further, the parameters in Eq.~(\ref{two}) are given by $g=h$,
$u=2$, and $a$ is given in Eq.~(\ref{const}). If $2\pi/N \ll h^q \ll 1$, the 
FS should indicate the location of the QCP if $q =1$, but not if $q \ge 2$. 

\begin{figure}[htb] \ig[width=3.4in]{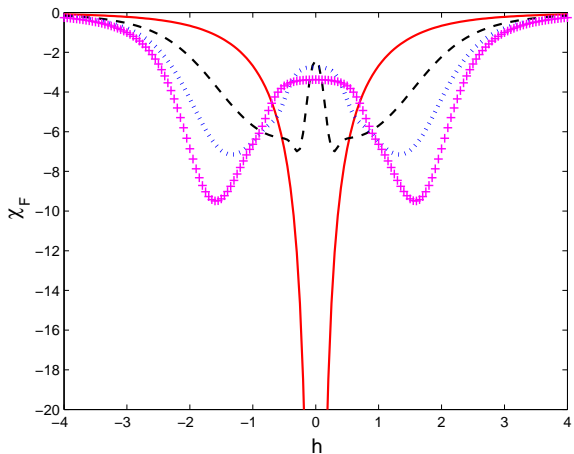}
\caption[]{(Color online) Plot of $\chi_F$ versus $h$ for $q=1, ~\phi=0$ 
(red solid), $q=2,~ \phi=\pi/4$ (black dashed), $q=3, ~\phi=0$ (blue dotted), 
and $q=4, ~\phi= \pi/8$ (magenta plus), with $N=240$ and $\theta =\pi/N$.} 
\label{fidfig1} \end{figure}

Fig.~\ref{fidfig1} confirms the above statement for $q=1,~2,~3$ and 4, $N=240$ 
and $\ta = \pi/N$. For each value of $q$ in that figure, we have chosen $\phi$
so as to maximize the gap $|\Delta|$ which is proportional to $\cos (q\phi)$
($\sin (q\phi)$) for $q$ odd (even). At $h=0$, there is a large peak in 
$-\chi_F$ for $q=1$, but not for $q=2, ~3$ and 4. For $q=1$, the peak value of 
$-\chi_F$ is found to be 450 at $h=0$ which lies far outside the range of 
Fig.~\ref{fidfig1}; we note that this peak value is consistent with 
Eq.~(\ref{fs4}) since $\nu =1$, $a=1$, and the Fermi velocity is $u=2$. 

In producing Fig.~\ref{fidfig1}, we have chosen $N$ to be a multiple of 4 
and $\ta = \pi/N$ for the following reason. As discussed in Eq.~(\ref{heff}),
the momentum $k$ introduced in 
Sec. II is actually the deviation from the momenta $\pm \pi/2$ for the model
discussed in Sec. III A. To prevent the expression in Eq.~(\ref{fs4}) from 
diverging in our model, we must ensure that $\pm \pi/2 + \ta + 2\pi n/N$ 
does not vanish for any value of $n$. This will be true if $N$ is a multiple
of 4 and $\ta= \pi/N$. (Alternatively, we could have chosen $N-2$ to be a
multiple of 4 and $\ta = 0$).

We observe in Fig.~\ref{fidfig1} that $\chi_F$ is not zero at $h=0$ for 
$q= 2, ~3$ and 4, in contrast to the result in Eq.~(\ref{fs4}). We can 
analytically compute $\chi_F (h=0)$ by using the following result from first 
order perturbation theory. If $H(\la) = H_0 + \la V$ is a many-body
Hamiltonian with eigenvalues $E_\al (\la)$ and eigenstates $\psi_\al (\la)$,
where $\al =0$ denotes the ground state, then the FS is given by \ct{grandi10}
\be \chi_F (\la) ~=~ - ~\frac{1}{2} ~\sum_{\al \ne 0} ~\frac{|\bra \psi_\al 
(\la) | V | \psi_0 (\la) \ket |^2}{(E_\al (\la) ~-~ E_0 (\la))^2}. \ee
We now apply this result to our model. At $h=0$
and $\ta = \pi/N$, the ground state of the total system is one in which the 
$N/2$ one-particles states with $k=-\pi/2, -\pi/2 + 2\pi/N, \cdots, \pi/2 - 
2\pi/N$ are filled and the remaining $N/2$ states are empty. Next, at $h=
dh$, the change in the ground state to first order in $dh$ is given by a
state in which a fermion has moved from a filled state at $k$ to an empty 
state at $k-\pi/q$ if $k$ lies in the range $[-\pi/2,-\pi/2+\pi/q]$, or from
$k$ to $k+\pi/q$ if $k$ lies in the range $[\pi/2-\pi/q,\pi/2]$. Using 
Eqs.~(\ref{det}-\ref{fs6}), we find that the FS at $h=0$ is given by 
\ba && \chi_F (h=0) \non \\
&=& - \sum_{k=-\pi/2}^{-\pi/2+\pi/q} ~\frac{1}{16 ~[\cos (k+\ta)
- \cos (k-\pi/q +\ta)]^2}, \non \\
&& \label{fs7} \ea
where $k$ goes in steps of $2\pi/N$, and we have used the fact that the 
contribution to $\chi_F$ from the range $[-\pi/2,-\pi/2+\pi/q]$ is equal to
the contribution from the range $[\pi/2-\pi/q,\pi/2]$. For large $N$, we can 
change the summation in Eq.~(\ref{fs7}) to an integral ($\sum_k \to \int dk 
(N/2\pi)$) and ignore $\ta$ to obtain
\be \chi_F (h=0) ~=~ - ~\frac{N}{32\pi \sin (\pi/q)}. \label{fs8} \ee
Note that this result does not depend on the phase $\phi$ in Eq.~(\ref{ham1}).
For $N=240$ and $q=2, ~3$ and 4, this gives $\chi_F (h=0) \simeq 2.39$, $2.76$
and $3.38$ respectively, which agree well with the values shown in 
Fig.~\ref{fidfig1}.

Finally we note in Fig.~\ref{fidfig1} that the FS for $q=2, ~3$ and 4 has a 
double peak 
structure; however, these peaks are unrelated to the QCP at $h=0$. We notice
that as $q$ increases, these peaks move towards $h = \pm 2$. We will see in 
Sec. IV that in the limit $q \to \infty$, there are QCPs at $h = \pm 2$ in 
addition to the QCP at $h=0$, and the double peaks are an indication of those
QCPs.

\subsection{Scaling of fidelity susceptibility with $\ta$}

In this section, we will show that by varying $\ta$, we can locate the QCP 
at $h=0$ even for $q \ge 2$. Further, the scaling of the FS with respect to 
$\ta$ allows us to find the value of the critical exponent $\nu$. At the end 
we will also discuss the scaling of the FS with $\phi$.

\begin{figure}[htb] \ig[width=3.4in]{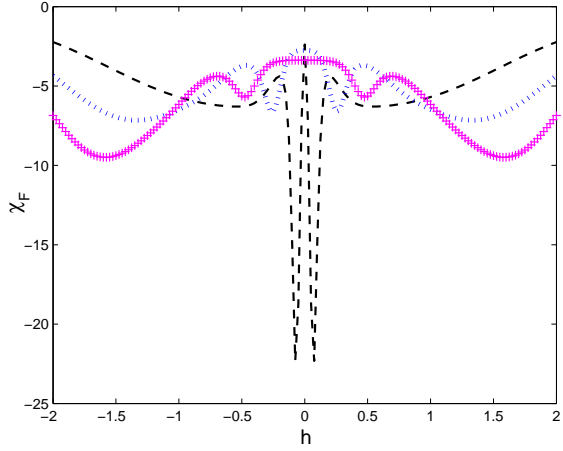}
\caption[]{(Color online) Plot of $\chi_F$ versus $h$ for $q=2,~ \phi=\pi/4$ 
(black dashed), $q=3, ~\phi=0$ (blue dotted), and $q=4, ~\phi= \pi/8$ (magenta
plus), with $N=240$ and $\theta =0.001$.} \label{fidfig2} \end{figure}

In Fig.~\ref{fidfig2}, we show the FS for $q=2, ~3$ and 4, $N=240$ and $\ta 
= 0.001$, which is much smaller than the value of $\ta = \pi/N$ chosen in 
Fig.~\ref{fidfig1}; the values of $\phi$ chosen for each $q$ are the same as
in Fig.~\ref{fidfig1}. We now see that additional double peaks have appeared
in the different curves which lie much closer to $h=0$, with the peak values 
decreasing and their distances from $h=0$ increasing as $q$ increases. We 
will now see that these new peaks are related to the QCP at $h=0$ and that 
their peak values and positions scale in accordance with Eq.~(\ref{scal2}); 
as $\ta \to 0$, the locations of the peaks approach the QCP.

\begin{figure}[htb] \ig[width=3.4in]{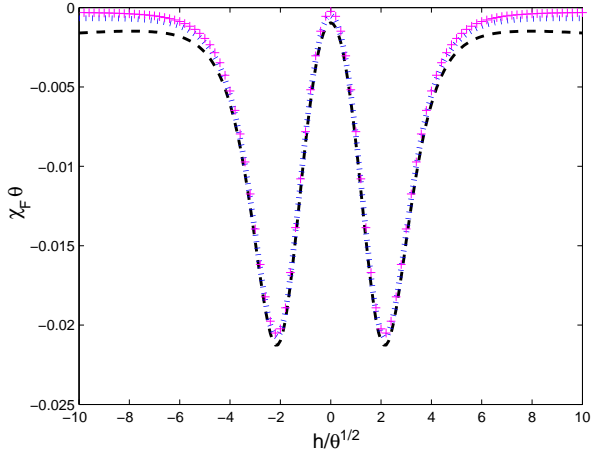}
\caption[]{(Color online) Plot of $\chi_F \ta$ versus $h/\ta^{1/2}$ for $\ta =
0.0004$ (black dashed), $\ta = 0.0002$ (blue dotted), and $\ta =0.0001$ 
(magenta plus), with $q=2$, $N=240$ and $\phi = \pi/4$.} \label{fidfig3} 
\end{figure}

\begin{figure}[htb] \ig[width=3.4in]{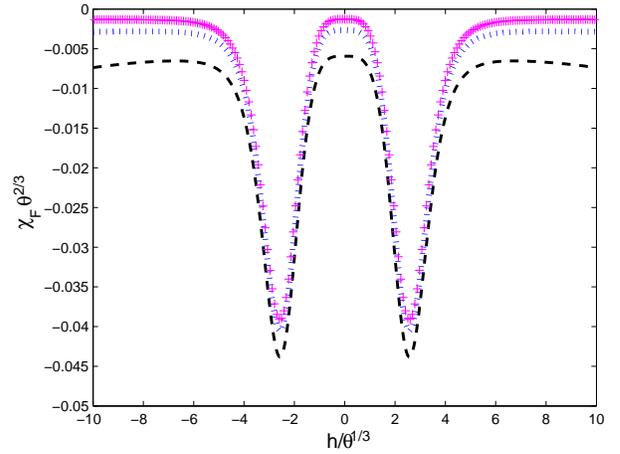}
\caption[]{(Color online) Plot of $\chi_F \ta^{2/3}$ versus $h/\ta^{1/3}$ for 
$\ta = 0.0001$ (black dashed), $\ta = 0.00003$ (blue dotted), and $\ta =
0.00001$ (magenta plus), with $q=3$, $N=240$ and $\phi = 0$.} \label{fidfig4} 
\end{figure}

In Fig.~\ref{fidfig3}, we show $\chi_F \ta^{2/q}$ versus $h/\ta^{1/q}$ for 
$q=2$, $N=240$ and $\phi = \pi/4$, with $\ta = 0.0004, ~0.0002$ and $0.0001$.
We see that the three curves fall on top of each other, thus confirming
the scaling form given in Eq.~(\ref{scal2}).
In Fig.~\ref{fidfig4}, we show $\chi_F \ta^{2/q}$ versus $h/\ta^{1/q}$ for 
$q=3$, $N=240$ and $\phi = 0$, with $\ta = 0.0001, ~0.00003$ and $0.00001$.
Once again the curves fall on top of each other.

The values of $\ta$ in Figs.~\ref{fidfig3} and \ref{fidfig4} indicate that as 
$q$ increases from 2 to 3, we have to go to smaller values of $\ta$ to see the
scaling form. The reason for this is as follows. We saw below Eq.~(\ref{scal1})
that the peak in the FS occurs at $g/ (u\ta)^{1/q} = [(q-1)/(q+1)]^{1/(2q)}$ 
when $g^q, \ta \ll 2\pi u/N$. In our model, $g=h$, $u=2$, and $a$ is given in 
Eq.~(\ref{const}). Putting these together, we find that the peak occurs at 
$\ta = (\sqrt{3}/8) ~h^2$ for $q=2$ and at $\ta = (\sqrt{2}/24) ~h^3$ for 
$q=3$. We thus see that for a small value of $h$, $\ta$ is smaller for $q=3$
compared to $q=2$. For $h=0.1$, for instance, the peak lies at $\ta \simeq 
0.0022$ for $q=2$ and at $\ta \simeq 0.000059$ for $q=3$.

\begin{figure}[htb] \ig[width=3.4in]{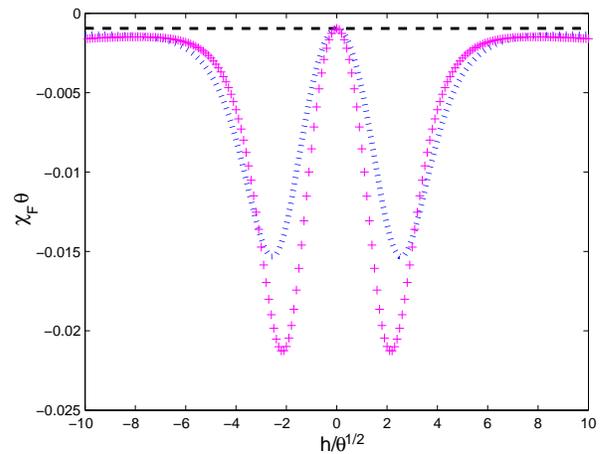}
\caption[]{(Color online) Plot of $\chi_F \ta$ versus $h/\ta^{1/2}$ for $\phi 
= 0$ (black dashed), $\phi = \pi/8$ (blue dotted), and $\phi =\pi/4$ (magenta 
plus), with $q=2$, $N=240$ and $\ta = 0.0004$.} \label{fidfig5} \end{figure}

\begin{figure}[htb] \ig[width=3.4in]{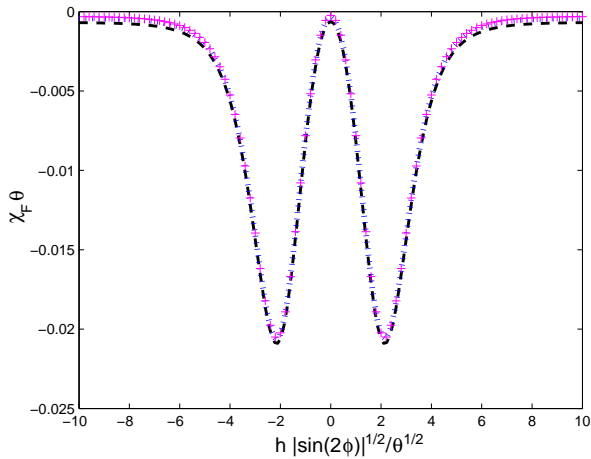}
\caption[]{(Color online) Plot of $\chi_F \ta$ versus $h|\sin (2\phi)|^{1/2}/
\ta^{1/2}$ for $\phi = \pi/16$ (black dashed), $\phi = \pi/8$ (blue dotted), 
and $\phi =\pi/4$ (magenta plus), with $q=2$, $N=240$ and $\ta = 0.0001$.} 
\label{fidfig6} \end{figure}

It is interesting to consider the dependence of the FS on the phase $\phi$.
At $h=0$, we saw that $\chi_F$ is given by Eq.~(\ref{fs8}) and is independent
of $\phi$. But $\chi_F$ does depend on $\phi$ away from $h=0$. 
Fig.~\ref{fidfig5} shows $\chi_F \ta$ versus $h/\ta^{1/2}$ for $q=2$, $N=240$ 
and $\ta = 0.0004$, with $\phi = 0, ~\pi/8$ and $\pi/4$. The three curves look
quite different; in fact, the curve for $\phi = 0$ hardly changes with $h/
\ta^{1/2}$ within the range shown in the figure. Note that $\phi =0$ is the 
value at which the perturbatively calculated gap $\Delta$ given by 
Eq.~(\ref{pert3}) vanishes for $q=2$ and is therefore independent of $h$ 
within the perturbative range. Hence $\chi_F$ is essentially independent of $h$
and is simply given by its value at $h=0$ if $\phi =0$. Using 
Eqs.~(\ref{scal1}) and (\ref{const}), we find that for $q=\nu=2$, $\chi_F \ta$
should be a function of $h|\sin (2\phi)|^{1/2}/\ta^{1/2}$. In 
Fig.~\ref{fidfig6}, we show $\chi_F \ta$ versus $h|\sin (2\phi)|^{1/2}/
\ta^{1/2}$ for $q=2$, $N=240$ and $\ta = 0.0001$, with $\phi = \pi/16, ~\pi/8$ 
and $\pi/4$. We see that the three curves fall on top of each other thus
confirming the presence of the factor of $|\sin (2\phi)|^{1/2}$ in the 
scaling function.

\subsection{Fidelity susceptibility in another model with $\nu =2$}

It turns out that there is another model in one dimension which has QCPs with 
$\nu =2$. This is an anisotropic spin-1/2 $XY$ spin chain in which the strength
of a transverse field alternates between $h+\de$ and $h-\de$ at odd and even 
sites \ct{deng,divakaran,perk,okamoto}. The Hamiltonian of the model is given 
by
\ba H &=& - ~\sum_n ~[~{\frac{(J_x + J_y)}{4}} ~(\si^x_n \si^x_{n+1} + 
\si^y_n \si^y_{n+1}) \non \\
& & ~~~~~~~~~~+~ {\frac{(J_x-J_y)}{4}} ~(\si^x_n \si^x_{n+1} - \si^y_n 
\si^y_{n+1}) \non \\
& & ~~~~~~~~~~+~ \frac{(h-(-1)^n\de )}{2} ~\si^z_n]. \ea
The spectrum of this model can be solved by carrying out the Jordan-Wigner 
transformation from spin-1/2's to spinless fermions at each site. (The system 
then decouples into a number of four-dimensional subsystems. We will not 
present the details here and refer the reader to Refs. \onlinecite{deng} and 
\onlinecite{divakaran}). Defining $J_x + J_y = J$ and $J_x - J_y = \ga$, we 
find that the model has four quantum critical lines given by $\de = \pm 
\sqrt{h^2 + \ga^2}$ and $h = \pm \sqrt{\de^2 + J^2}$. If $\de = \pm \ga$ is 
held fixed, 
there are QCPs at $h=0$ which lie on the critical lines $\de = \pm \sqrt{h^2 
+ \ga^2}$. Alternatively, if $h = \pm J$ is held fixed, there are QCPs at 
$\de = 0$ which lie on the critical lines $h = \sqrt{\de^2 + J^2}$. All these 
QCPs have $z=1$ and $\nu =2$. 

\begin{figure}[htb] \ig[width=3.4in]{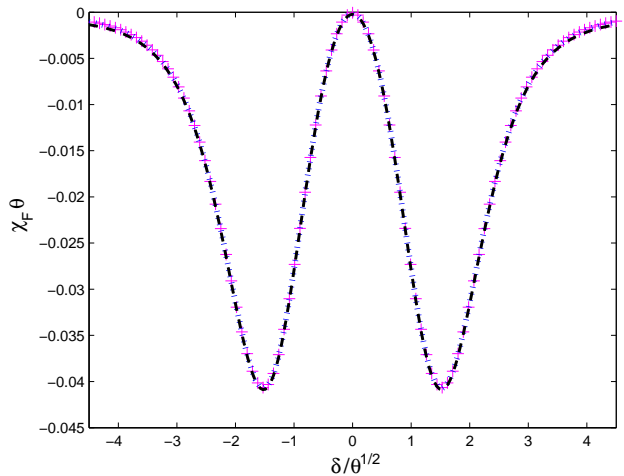}
\caption[]{(Color online) Plot of $\chi_F \ta$ versus $\de/\ta^{1/2}$ for $\ta 
= 0.0004$ (black dashed), $\ta = 0.0002$ (blue dotted), and $\ta =0.0001$ 
(magenta plus) for the anisotropic spin-1/2 chain with $N=240$.} 
\label{fidfig7} \end{figure}

We have numerically studied the FS of this model after introducing a twist
$\ta$ in the Jordan-Wigner fermionic Hamiltonian. Holding $J=2$, $\ga =1$ and 
$h=2$ fixed, we varied $\de$ to go through the QCP lying at $\de = 0$.
Fig.~\ref{fidfig7} shows the scaling of $\chi_F \ta$ versus $\de/\ta^{1/2}$
for $N=240$ and $\ta = 0.0004$, $0.0002$ and $0.0001$. The scaling near the 
QCP at $\de=0$ is very similar to what is seen in Fig.~\ref{fidfig3}, 
just as one would expect from Eq.~(\ref{scal2}) for a system with $\nu =2$. 
This model therefore confirms that the introduction of a twist angle can 
enable the FS to detect a QCP with $\nu =2$.

\section{Fidelity susceptibility for very long periods}

If the quantity $\pi/q$ in Eq.~(\ref{ham1}) is replaced by $\pi$ times an 
irrational number (which can be approximated by rational numbers with 
increasingly large denominators), we obtain a quasiperiodic system 
\ct{azbel,hsu,aubry,ober,ost,sok,delyon,sun,wieg,aulbach,modugno}. It is 
known that such a system has a metal-insulator transition at $h=\pm 2$; the 
nature of the eigenstate at zero energy changes from extended (metallic) to 
localized (insulating) on crossing these QCPs \ct{sok,delyon,aulbach,modugno}.
The FS has been used to detect QCPs in quasiperiodic systems \ct{guang} as 
well as in disordered systems \ct{cestari}.

In this section, we will study the FS of the model defined in Eq.~(\ref{ham1})
as $q$ becomes very large. In that limit, we find numerically that the FS has
increasingly large peaks at $h = \pm 2$; we have seen precursors of these
peaks in Fig. 1 for $q=2,~3$ and 4. As far as we know, these QCPs have not 
been reported earlier. Our results indicate that the metal-insulator transition
also seems to occur in our model, although $1/q$ approaches zero rather than
an irrational number as $q \to \infty$. 

\begin{figure}[htb] \ig[width=3.4in]{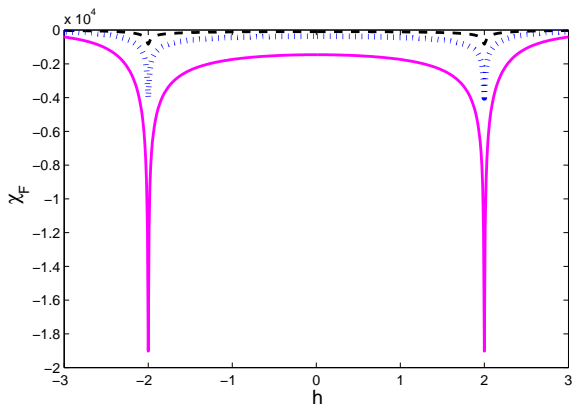}
\caption[]{(Color online) Plot of $\chi_F$ versus $h$ for $q=120$ (black 
dashed), $q=240$ (blue dotted), and $q=480$ (magenta solid), with $N=2q$, $\ta 
= \pi/N$ and $\phi=\pi/N$ in each case.} \label{fidfig8} \end{figure}

\begin{figure}[htb] \ig[width=3.4in]{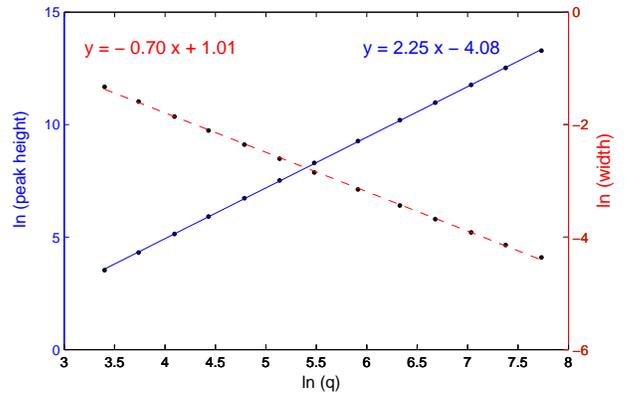}
\caption[]{(Color online) Log-log plot of the peak value (blue solid, $y$-axis
on left) and the full width at half maximum (red dashed, $y$-axis on right) 
versus $q$ for the peak in $\chi_F$ at $h=2$.} \label{fidfig9} \end{figure}

\begin{figure}[htb] \ig[width=3.4in]{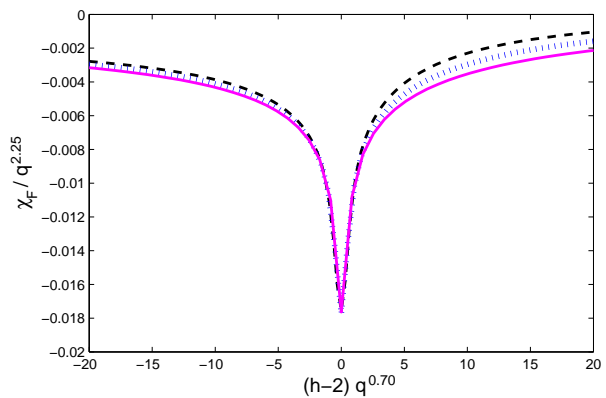}
\caption[]{(Color online) Plot of $\chi_F /q^{2.25}$ versus $(h-2)q^{0.70}$ 
for $q=120$ (black dashed), $q=240$ (blue dotted), and $q=480$ (magenta solid),
with $N=2q$, $\ta = \pi/N$ and $\phi=\pi/N$ in each case.} \label{fidfig10} 
\end{figure}

In Fig.~\ref{fidfig8}, we show the FS for $q=120, ~240$ and 480, with $N=2q$, 
$\ta = \pi/N$ and $\phi=\pi/N$ in each case. (Since $N$ must be a multiple 
of $2q$ and $q$ is quite large, we are only presenting the results for $N=2q$ 
here. However, we have checked for $q = 60$ that our results do not change if 
we take $N$ to be a higher multiple of $2q$. Further, we have fixed the values
of $\ta$ and $\phi$ in terms of $N$, and will not study here how the FS varies
with them). We observe prominent peaks in the FS at $h=\pm 2$. 
Fig.~\ref{fidfig9} shows a log-log plot of the peak value and the full width 
at half maximum versus $q$ for the peak in $\chi_F$ at $h=2$, for a range
of values of $q$ from 30 to 2280. We find that the peak value of $\chi_F$ 
scales as $q^{2.25}$ while the full width at half maximum scales as 
$1/q^{0.70}$. In Fig.~\ref{fidfig10}, we show the scaling near $h=2$ by 
plotting $\chi_F /q^{2.25}$ versus $(h-2)q^{0.70}$ for $q=120, ~240$ and 480; 
the curves fall on top of each other. Interestingly, we see that the curves 
are somewhat asymmetric about $h=2$.

We have also studied the energy gap and the nature of the wave function at 
zero energy, i.e., the Fermi energy. Upon extrapolating to the thermodynamic 
limit $N \to \infty$, we find that the gap is zero even if we are away from 
$h=\pm 2$; hence these QCPs are different from the ones discussed in the 
earlier sections where the gap scales as $|h-h_c|^\nu$ and is therefore 
non-zero for $h \ne h_c$ even if $N \to \infty$ (here $h_c$ denotes the 
location of the QCP). The QCPs at $h=\pm 2$ are not characterized by the gap 
going to zero but rather by a change in the nature of the wave function at 
zero energy.

We can understand why the behavior of the system changes at $h = \pm 2$ by 
using a continuum theory to study the properties of Eq.~(\ref{ham1}) for a 
state whose energy lies at zero, i.e., at the Fermi energy. (We will 
assume here that the limit 
$N \to \infty$ has been taken). A continuum theory is justified if $q$ 
is very large since the chemical potential then varies on a length scale
which is much longer than the lattice spacing $a$. We can remove the twist 
angle $\ta$, by performing the phase transformation $c_n \to e^{-in\ta}
c_n$ in Eq.~(\ref{ham1})), if we are only interested in the behavior of a 
state in a local region of the system. Setting $J=1$, the equation of motion 
following from Eq.~(\ref{ham1}) is
\be - ~(c_{n+1} ~+~ c_{n-1}) ~-~ h \cos (\frac{\pi n}{q} + \phi) ~c_n ~=~ E ~
c_n \label{eom1} \ee
for a state with energy $E$. If $c_n$ varies slowly with $n$, we can write 
$2 c_n - c_{n+1} - c_{n-1} = - d^2 c(x)/dx^2$, where $x=na$ (we will set $a=
1$). Assuming $h>0$, we redefine $n$ as $n-n_0$ where $n_0$ is an integer 
chosen in such a way that $\pi n_0 / q + \phi$ is as close to $\pi$ as 
possible; hence $-h \cos (\pi n_0 / q + \phi)$ is close to its maximum value 
of $h$. (We can choose $n_0$ in an infinite number of ways; the various 
choices differ from each other by multiples of $2q$). We then find that
Eq.~(\ref{eom1}) can be written as a differential equation for $c(x)$,
\be -~\frac{d^2 c}{dx^2} ~+~ [h \cos (\frac{\pi x}{q}) ~-~ 2] ~c ~=~ E ~c. 
\label{eom2} \ee
Eq.~(\ref{eom2}) describes a particle moving in a periodic potential whose
maximum value is $h-2$. The potential has an infinite number of wells, each
extending from $n_0$ to $n_0 + 2q$. If $h<2$, a state with $E=0$ lies above 
the maximum of the potential; in that case the particle can move classically 
between the different wells, and the wave function will be extended 
throughout the system. If $h>2$, a particle with $E=0$ is classically confined
to one particular well of the periodic potential and can only go to other 
wells by quantum mechanical tunneling. If $h$ is slightly greater than 2, a 
WKB approximation shows that the tunneling probability between two neighboring 
wells is proportional to $\exp [-(h-2)q]$. If $(h-2)q >> 1$, the 
tunneling probability is extremely small, and the wave function is localized 
within a single well. We thus see that in the limit $q \to \infty$, $h=2$ 
marks a transition between extended and localized wave functions for the 
state with zero energy.

\section{Conclusions}

To summarize, we have shown that in a one-dimensional model which has a QCP 
with a correlation length exponent $\nu \ge 2$, the introduction of a twist
angle $\ta$ enables us to use the fidelity susceptibility to determine the 
location of the QCP. Namely, if the Hamiltonian has a parameter $g$ such
that the QCP lies at $g=0$, the twist allows us to bring the energy of
a particular state close to zero. If $\ta \ll 2\pi/N$, the FS scales as
$\chi_F \sim \ta^{-2/\nu} f(g/\ta^{1/\nu})$ which makes the QCP clearly
visible if $\chi_F$ is plotted versus $g$. A twisted boundary condition 
therefore provides a powerful tool for locating a QCP which may be difficult 
to find in any other way. We have argued that this technique may also be
useful for finding QCPs in higher dimensional models.

The specific model that we have used to demonstrate this idea is a 
tight-binding model of spinless fermions with a periodic chemical potential 
with amplitude $h$ and period $2q$, where $q$ is an integer. We have studied 
in detail the QCP lying at $h=0$; at this point, the dynamical critical 
exponent is given by $z=1$ while $\nu = q$. This makes this model specially 
useful for testing what happens for different values of $\nu$. We have shown 
analytically, using perturbation theory about $h=0$ and the decoupling of the 
system into a number of $(2q)$-dimensional subsystems, that the fidelity 
susceptibility scales as $\chi_F \sim \ta^{-2/\nu} f(h/\ta^{1/\nu})$ if 
$\ta$ is sufficiently small; we have verified this scaling numerically for 
small values of $q$. Although we have not presented the details here, we have 
confirmed that a similar scaling relation holds in another model which has 
QCPs with $\nu =2$. This is an anisotropic spin-1/2 $XY$ chain with 
transverse fields which has alternating strengths on odd and even sites.

In the last part of the paper, we have considered what happens in our model
when $q$ becomes very large. We find that some additional QCPs appear at $h= 
\pm 2$ in that limit, and the FS is clearly able to detect these QCPs. To
the best of our knowledge, these QCPs have not been reported before. We have 
studied the power laws associated with the peak value and the width of the FS 
as a function of $q$; we find non-trivial powers of $2.25$ for the peak value 
and $0.70$ for the width. The energy gap 
between the ground state and the first excited state is zero in the 
thermodynamic limit $N \to \infty$ over a finite range of values of $h$ 
around these QCPs. This makes it difficult to define the exponent $\nu$ at 
these QCPs (unlike the QCP lying at $h=0$ where we know that $\nu = q$).
Using a continuum theory, we have argued that these QCPs are characterized 
by a change in the nature of the wave function of a particle at the Fermi 
energy from extended to localized. In the future it would be interesting to 
develop a more detailed understanding of these QCPs which may shed
some light on the non-trivial power laws that we have found.

The analysis in this paper is expected to be valid for any one-dimensional
system which reduces to a theory of non-interacting fermions in the 
low-energy limit, i.e., close to the Fermi energy. Near a QCP with $z=1$ and 
an arbitrary value of the correlation length exponent $\nu$, the modes near 
the critical momenta ($= \pm \pi/2$ for a half-filled system) can be described
by $2 \times 2$ Hamiltonians as in Eqs. (\ref{two}) and (\ref{heff}). From
this we can deduce the scaling of the fidelity susceptibility with $g$ and 
$\ta$. We also argued at the end of Sec. II A that the analysis can be 
generalized to the case of $z$ not equal to 1 and in Sec. II B to theories 
of non-interacting fermions in higher dimensions.

The case of interacting systems is more complicated. In one dimension, such 
systems are typically described by Tomonaga-Luttinger liquid theory which has 
$z=1$. In a recent paper \ct{manisha}, some of us studied quenching in an 
interacting system which is closely related to the model considered in this 
paper. We argued there that the effect of interactions is to change the value 
of $\nu$ from $q$ to $q/(2-K)$ where the parameter $K$ depends on the strength
of the interactions, provided that $K$ lies in the range $1/2 < K < 2$. (A 
non-interacting system has $K=1$, and we then recover the result $\nu = q$). 
We expect that a similar result would hold for the fidelity susceptibility of
interacting systems; this may be an interesting question for future 
investigation.

Finally, we note that the twisted boundary condition has been introduced
in this paper as a mathematical device to produce a non-trivial scaling of 
the fidelity susceptibility which can provide information about the value of 
$\nu$. However, such a boundary condition also admits an interesting physical
interpretation. A one-dimensional system with periodic boundary conditions is 
the same as a circle. Imposing twisted boundary conditions in such a system 
is equivalent to assigning a charge to the particles and passing a 
magnetic flux through the middle of the circle so that the Aharonov-Bohm phase
(the product of the charge and the magnetic flux in some appropriate units) 
is equal to the twist angle $N\ta$. By introducing a twist we are therefore 
effectively studying the effect of a magnetic flux on the fidelity 
susceptibility of a system of charged particles moving on a circle.

\acknowledgments
For financial support, M.T. and A.D. thank CSIR, India and D.S. thanks DST, 
India for Project No. SR/S2/JCB-44/2010.

\end{document}